\documentclass[%
 reprint,
superscriptaddress,
amsmath,amssymb,
aps,
floatfix,
]{revtex4-2}

\usepackage{graphicx}
\usepackage{dcolumn}
\usepackage{bm}
\usepackage{booktabs}
\usepackage{multirow}
\usepackage{siunitx}
\DeclareSIUnit\angstrom{\text{\AA}}
\usepackage{float}
\usepackage[normalem]{ulem}

\begin{document}
 
\preprint{APS/123-QED}

\title{Skyrmion Alignment and Pinning Effects in a Disordered Multi-Phase Skyrmion Material Co$_{8}$Zn$_{8}$Mn$_{4}$}

\author{M. E. Henderson}
\email{mehenderson@uwaterloo.ca}
\affiliation{Institute for Quantum Computing, University of Waterloo, Waterloo, ON, Canada, N2L3G1}
\affiliation{Department of Physics \& Astronomy, University of Waterloo, Waterloo, ON, Canada, N2L3G1}

\author{M. Bleuel}
\affiliation{National Institute of Standards and Technology, Gaithersburg, Maryland 20899, USA}
\affiliation{Department of Materials Science and Engineering, University of Maryland, College Park, MD  20742-2115, USA}

\author{J. Beare}
\affiliation{Department of Physics and Astronomy, McMaster University, Hamilton, ON, Canada, L8S 4M1}

\author{D. G. Cory}
\affiliation{Institute for Quantum Computing, University of Waterloo, Waterloo, ON, Canada, N2L3G1}
\affiliation{Department of Chemistry, University of Waterloo, Waterloo, ON, Canada, N2L3G1}

\author{B. Heacock}
\affiliation{National Institute of Standards and Technology, Gaithersburg, Maryland 20899, USA}

\author{M. G. Huber}
\affiliation{National Institute of Standards and Technology, Gaithersburg, Maryland 20899, USA}
\author{G. M. Luke}

\affiliation{Department of Physics and Astronomy, McMaster University, Hamilton, ON, Canada, L8S 4M1}
\affiliation{Brockhouse Institute for Materials Research, Hamilton, ON, Canada, L8S 4M1}

\author{M. Pula}
\affiliation{Department of Physics and Astronomy, McMaster University, Hamilton, ON, Canada, L8S 4M1}

\author{D. Sarenac}
\affiliation{Institute for Quantum Computing, University of Waterloo, Waterloo, ON, Canada, N2L3G1}

\author{S. Sharma}
\affiliation{Department of Physics and Astronomy, McMaster University, Hamilton, ON, Canada, L8S 4M1}

\author{E. M. Smith}
\affiliation{Department of Physics and Astronomy, McMaster University, Hamilton, ON, Canada, L8S 4M1}

\author{K. Zhernenkov}
\affiliation{Institute for Quantum Computing, University of Waterloo, Waterloo, ON, Canada, N2L3G1}
\affiliation{J\"ulich Centre for Neutron Science at Heinz Maier-Leibnitz Zentrum, Forschungszentrum J\"ulich GmbH, 85748 Garching, Germany}

\author{D. A. Pushin}
\email{dmitry.pushin@uwaterloo.ca}%
\affiliation{Institute for Quantum Computing, University of Waterloo, Waterloo, ON, Canada, N2L3G1}
\affiliation{Department of Physics \& Astronomy, University of Waterloo, Waterloo, ON, Canada, N2L3G1}

\date{\today}
\begin{abstract}

Underlying disorder in skyrmion materials may both inhibit and facilitate skyrmion reorientations and changes in topology. The identification of these disorder-induced topologically active regimes is critical to realizing robust skyrmion spintronic implementations, yet few studies exist for disordered bulk samples. Here, we employ small-angle neutron scattering (SANS) and micromagnetic simulations to examine the influence of skyrmion order on skyrmion lattice formation, transition, and reorientation dynamics across the phase space of a disordered polycrystalline Co$_{8}$Zn$_{8}$Mn$_{4}$ bulk sample. Our measurements reveal a new disordered-to-ordered skyrmion square lattice transition pathway characterized by the novel promotion of four-fold order in SANS and accompanied by a change in topology of the system, reinforced through micromagnetic simulations. Pinning responses are observed to dominate skyrmion dynamics in the metastable triangular lattice phase, enhancing skyrmion stabilization through a remarkable and previously undetected skyrmion memory effect which reproduces previous ordering processes and persists in zero field. These results uncover the cooperative interplay of anisotropy and disorder in skyrmion formation and restructuring dynamics, establishing new tunable pathways for skyrmion manipulation.
   
\end{abstract}

\maketitle

\section{\label{sec:level1}Introduction\protect\\}

Magnetic skyrmions represent localized spin configurations, where the spins wrap the entire unit sphere. This countable particle-like property, combined with electric controllability and unique transport phenomena via the quantum spin Berry phase mechanism, makes magnetic skyrmions attractive candidates for future non-volatile high density magnetic memory devices and logic computing applications \cite{Zhang2015magnetic, Wiesendanger2016nanoscale,Sampaio2013nucleation, Romming2013writing}. Magnetic skyrmions may form in noncentrosymmetric materials with competing exchange interactions which stabilize helical ground-states \cite{muhlbauer2009skyrmion,Yu2011near,Munzer2010skyrmion,Shibata2013towards}, or in centrosymmetric systems from additional magnetic contributions such as geometric frustration or four-spin interactions mediated by itinerant electrons \cite{geo, itinerent}. In the former, spin-orbit coupling generates an interfacial Dzyaloshinskii–Moriya interaction (DMI) at hetero-interfaces or bulk DMI in bulk materials, resulting from broken inversion symmetry.  Accordingly, skyrmions have been detected in ferromagnetic ultrathin multilayer films \cite{Heinze2011spontaneous,Romming2013writing,Dupe2014tailoring,thinfilm}, and chiral bulk magnets \cite{muhlbauer2009skyrmion,Wilhelm2011precursor,Munzer2010skyrmion,Seki2012observation,roomtemp,Kezsmarki2015neel}. In such systems, the intricate interplay of DMI with the exchange interaction, Zeeman energy, and magnetocrystalline anisotropies cooperatively determines the magnetic structures, their spin propagation vectors and periodicity \cite{Butch}. In zero magnetic field, the competition between the ferromagnetic exchange interaction and DMI stabilizes a helical state; under a magnetic field, the Zeeman energy stabilizes a skyrmion state, with the field direction typically defining the orientation of the skyrmion tubes. Magnetocrystalline anisotropy then helps determine the orientation of the skyrmion lattice itself.

Bulk skyrmion states have been realized over a broad temperature-magnetic field phase space, with stabilization mechanisms ranging from thermal fluctuations just below $T_{c}$ in thermodynamic equilibrium states \cite{roomtemp}, to quenching processes via rapid field-cooling (FC) procedures in metastable states \cite{glass}. Such metastable skyrmion phases have been shown to exhibit disorder-dependent cooling rates (i.e., dependent on doping, vacancies, etc, present in the system) \cite{robust,glass, kagawa2017current, doping, metastable,milde2013unwinding,memory}, with a unique hierarchy of interaction terms determining unconventional skyrmion formation and stabilization energetics \cite{roadmap}. One notable example is the anisotropy-stabilized low temperature skyrmion lattices realized in the bulk cubic helimagnet, Cu$_{2}$OSeO$_{3}$, through competition between anisotropic exchange and cubic anisotropy \cite{memory}. Additionally, studies of disordered skyrmion systems with weak magnetocrystalline anisotropy have demonstrated the precipitation of skyrmion order in triangular lattice phases through sequences which rock the external magnetic field relative to the sample \cite{dustin}. As such, examining the role of disorder and anisotropy in skyrmion systems \cite{Butch,doping} serves as an excellent testbed for realizing new pathways and mechanisms which stabilize metastable phases and facilitate structural lattice transitions.

 The chemically doped Co$_{x}$Zn$_{y}$Mn$_{20-x-y}$ compositional series presents a unique platform to investigate skyrmion behaviour, owing to it's interplay of magnetic anisotropy, site disorder, and frustration, which generates diverse topological phases and lattice forms \cite{metastable,robust,yu2018transformation, roomtemp}. Of particular interest are the thermal equilibrium and metastable phases, which host skyrmion lattices of both triangular and square forms. Random site occupancies of Co and Mn atoms on the 8c site, and Co, Zn, and Mn atoms on the 12d site introduces site-disorder which stabilizes a high density of defects \cite{spin,metastable}, enabling quenching to triangular lattice metastable phases with moderate cooling rates \cite{robust,metastable,elementukleev2019,deformationMorikawa2017}. Upon further cooling, the interaction energy hierarchy shifts: magnetocrystalline anisotropy increases \cite{prebinger2021vital,metastable} and the development of antiferromagnetic correlations of the Mn spins decrease the ratio of the ferromagnetic exchange to DM interaction \cite{elementukleev2019,metastable,spin,deformationMorikawa2017}. Together, these two actions drive a large increase in q, which triggers a triangular-to-square lattice transition, where anisotropy determines the directionality and type of distortion \cite{robust,metastable,elementukleev2019,deformationMorikawa2017}.

 We employ detailed SANS measurements on a bulk polycrystalline disordered sample of Co$_{8}$Zn$_{8}$Mn$_{4}$, previously reported in \cite{henderson2020characterization}, to investigate skyrmion formation, reorientation, and pinning dynamics as a function of skyrmion order across multiple phases. Using a  skyrmion-ordering sequence in which the sample was effectively rocked in the magnetic field \cite{dustin}, we were able to generate varying levels of skyrmion order in the thermal equilibrium phase, from which we investigated the triangular and square metastable transitions and phases. Rocking scans were also performed with the sample and field rotated coincident about the neutron beam over a 5.8 degree total angular range, limited by the bore of the magnet. Annular averages of the primary scattering rings in the SANS data are taken to highlight the types of skyrmion lattice symmetries present and quantify the degree of skyrmion order. When entered from the disordered thermal equilibrium phase, the square lattice exhibited commensurate four-fold order; the appreciation in skyrmion order during the transition suggests improved skyrmion alignment and development through nucleation and elongation along distinguished directions as anisotropy increases, reinforcing structural lattice transition mechanisms put forth in \cite{skyrmion_frac}. This is reinforced through micromagnetic simulations, which suggest a novel disordered-to-ordered square lattice transition pathway via a change in topology of the system. Saturation to-and-from the field-polarized ferromagnetic state, within the triangular lattice metastable temperature window, revealed a skyrmion memory effect, reproducing previous ordering processes and persisting even in zero field. Skyrmion ordering dynamics for different magnetic fields in this state demonstrated limited skyrmion reorientations/development, with previous pinning sites dominating skyrmion orientations and positions. Our findings elucidate the nature of skyrmion formation, lattice transitions, and stabilization mechanisms across skyrmion phase space.

\section{\label{sec:level2}Results\protect\\}

\subsection{\label{sec:level1}Skyrmion ordering in the thermal equilibrium phase  \protect\\}

The experimental setup for the SANS experiments is shown in Fig. 1a. The SANS pattern for the initial helical state was that of four smeared magnetic satellites atop a circular ring, indicative of multi-domain single q-helical structures (Fig. 1b) \cite{roomtemp}. After FC through the ferromagnetic phase, from 420 K in a field of 250 G, a ring developed at 310 K (Fig. 1c).  The absence of the signature triangular lattice skyrmion hexagonal pattern is most likely a combination of the chemically disordered and above-room temperature nature of the phase/material. 

\begin{figure}[ht]
\includegraphics[width = \columnwidth]{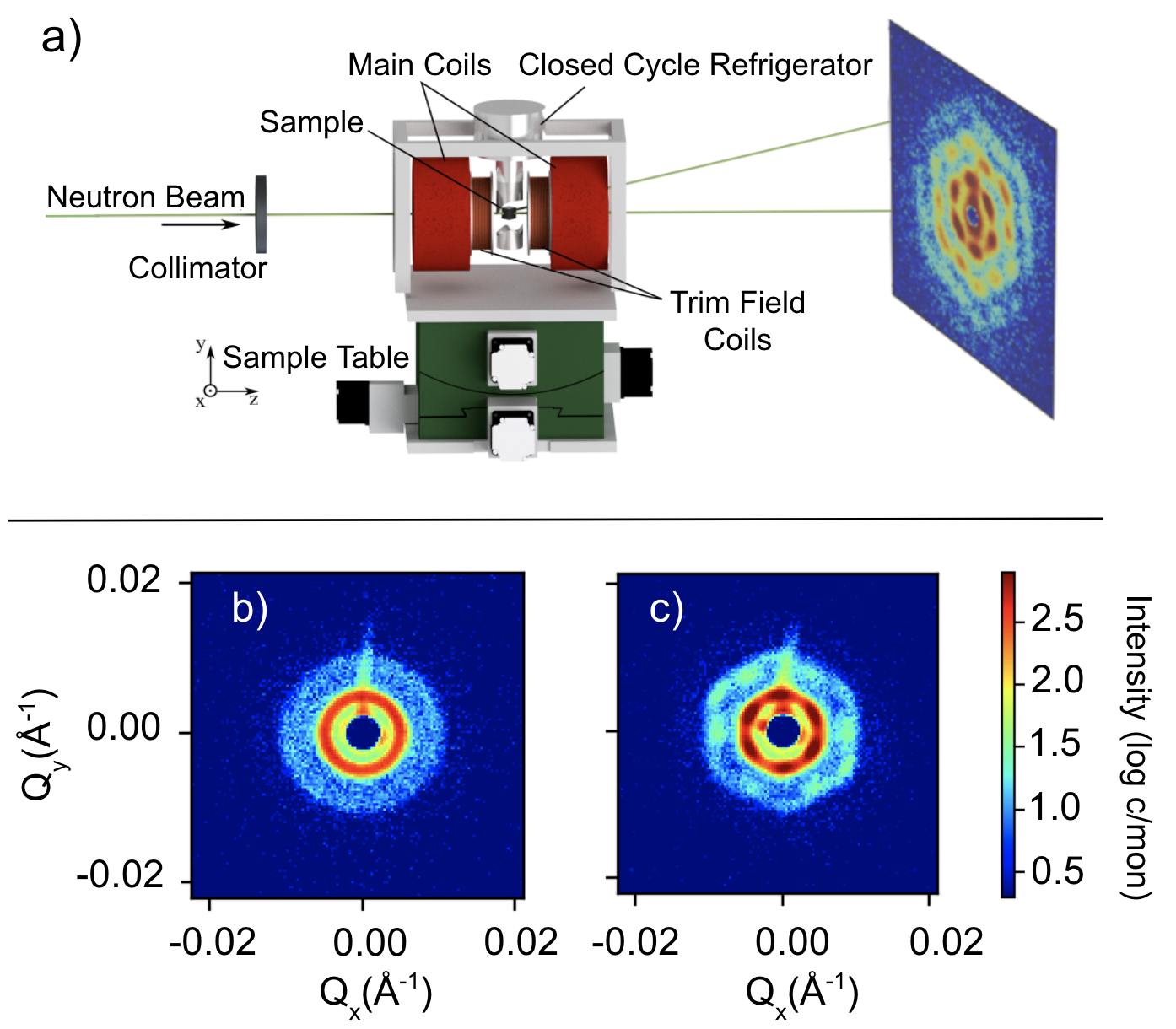}
\caption{\label{fig:realX} Schematic of the experimental SANS setup (a). The neutron propagation direction and magnetic field direction are parallel (along z). For the ordering sequence the sample was rotated symmetrically about the y-axis (in the xz-plane), with the magnetic field held fixed in the z-direction. SANS images for b) disordered helical state at 290 K in 0 G, and c) initially disordered thermal equilibrium  phase (D-Skx) at 310 K in 250 G, both exhibiting magnetic scattering at approximately the same $q_{0}$ of 0.0052 \AA $^{-1}$ (reproduced from \cite{henderson2020characterization}). Note that the presence of a flare in both images is caused by scattering off of the samples surface due to a neutron beam size exceeding the sample size. Both images are normalized to a fixed number of standard monitor counts to enable direct comparison between images, with the same scale for the color plots. Note the colorbar is a log-scale of intensity. }
\end{figure} 

\begin{figure}
\includegraphics[width=\columnwidth]{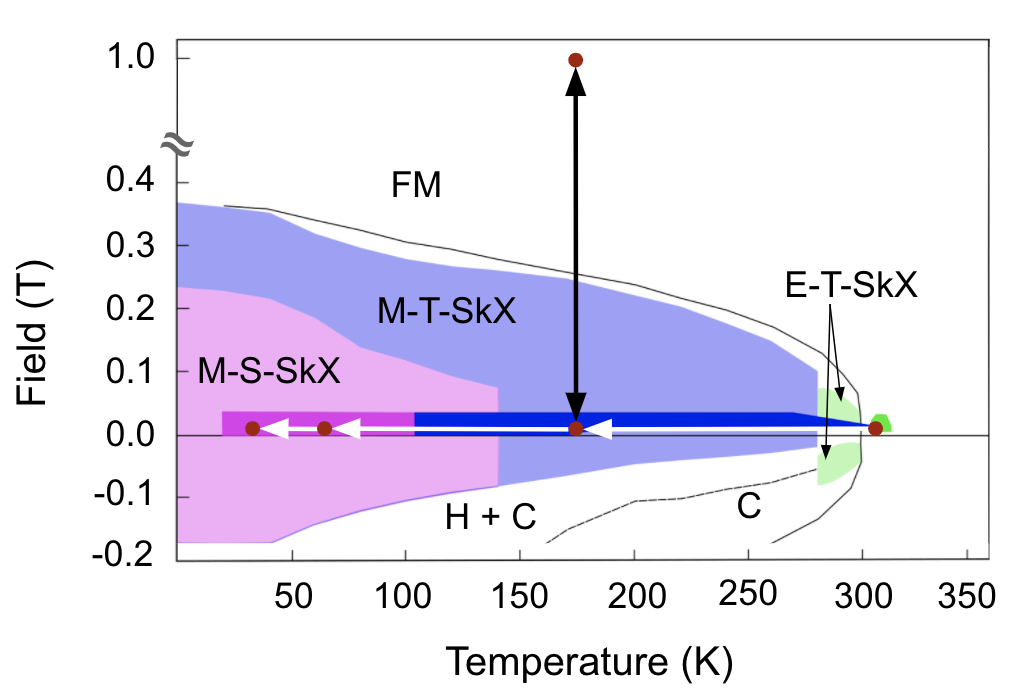}
\caption{Known phase diagram of  Co$_{8}$Zn$_{8}$Mn$_{4}$ material (lightly shaded regions) compiled from \cite{metastable}, with region of phase space which we have sampled through AC and SANS measurements (both in this paper and in \cite{henderson2020characterization}) illustrated by darkly shaded regions. Red dots illustrate measurement points of SANS images in the paper. White and black arrows indicating the field-cooling procedures and memory test performed in the experiments, respectively. Field-cooling procedures were performed over a magnetic field range spanning 50 G to 400 G. Green regions correspond to the thermal equilibrium skyrmion phase (E-T-SkX). Note: for our material, this skyrmion window is shifted up by approximately 10 K as compared to \cite{metastable}. Purple and pink regions correspond to metastable triangular lattice (M-T-SkX) and square lattice phases (M-S-SkX), respectively. Helical, conical, and ferromagnetic states are indicated by H, C, and FM, respectively.}\end{figure}

When referring to the levels of disorder in the material, there are two principle length scales we may address: the exchange interaction field (i.e the combination of ferromagnetic exchange (A), DMI (D), and anisotropy from site-to-site), and the periodicity of the spin-texture (as determined on average from the ratio of A/D). For this material, the exchange length is approximately 11 nm as defined in \cite{aboexchange2013} using material parameters from \cite{elementukleev2019}. The crystal mosaicity was previously reported in \cite{henderson2020characterization} as consisting of approximately 2-3 grains, with the crystal cut such that the (100) direction of the dominant grain is along the front face of the rectangular prism. Systematically scanning and slicing of the material using Backscatter X-ray Laue diffraction provided estimations of sample mosaicity consisting of 2-3 grains, one occupying a majority volume fraction, with low-angle grain boundaries. In order to produce a ring-like scattering image, one would need a large collection of slightly misoriented grains---assuming the grains are composed of idealized skyrmion domains which exhibit perfect triangular packing of the skyrmions (i.e., long transverse correlation lengths) and therefore minimal azimuthal smearing in their diffraction peaks. Each grain would then precipitate a sixfold pattern, with their superposition smearing the individual hexagonal patterns into a ring. In the event that the sample was comprised of a few grains, large-angle grain boundaries would be required to sufficiently rotate the hexagonal scattering patterns enough to smear their collective scattering pattern into a ring. Given the reported crystallinity of the sample, these two cases are unlikely to be responsible for the observed disorder. We will therefore concern ourselves primarily with the two length scales of disorder mentioned earlier, and take note that grain boundaries may act as sources of disorder, interrupting skyrmion long-range ordering in a non-linear manner \cite{MnSigrain}. While this effect is not quantifiable given only a single polycrystalline sample and the compounding influence of site-disorder, sample polycrystallinity may become more relevant when discussing defect-related pinning in the metastable phase.

\begin{figure*}\center
\includegraphics[width = \textwidth]{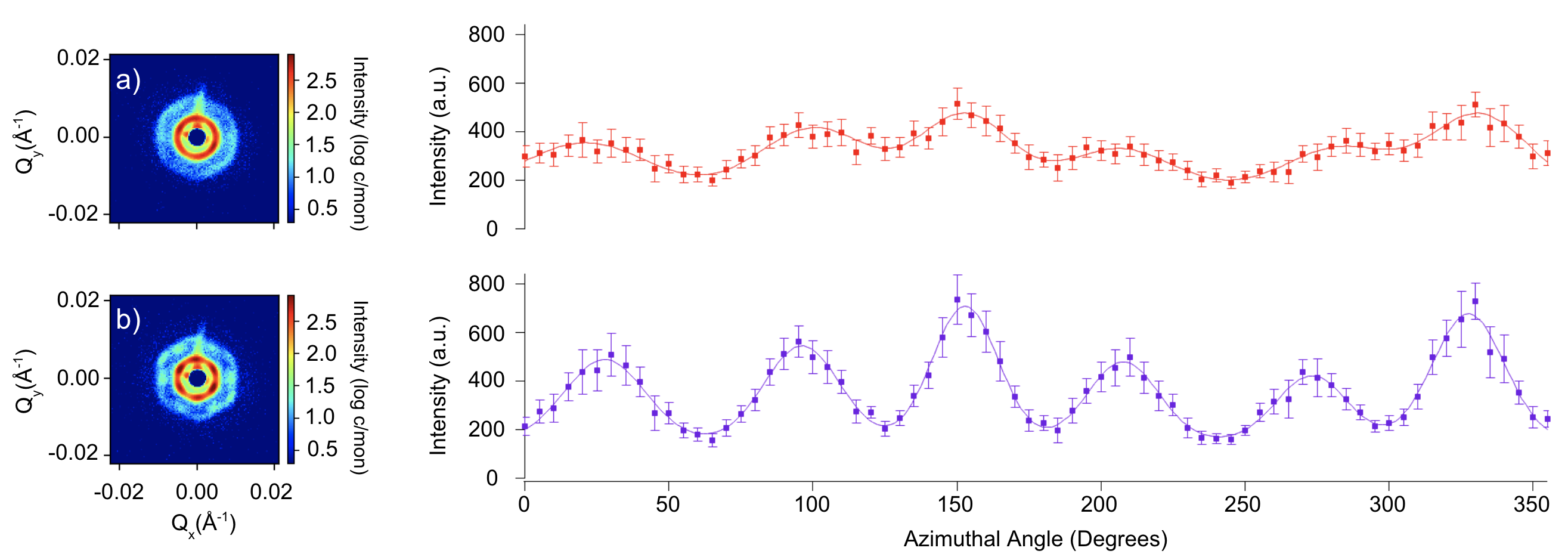}
\caption{ SANS measurements as a function of magnetic field skyrmion-ordering sequence. Diffraction patterns and their corresponding annular averages of first-order scattering rings after 10 (a), and 40 (b) rotations of the ordering sequence at 310 K in a magnetic field of 250 Oe. Solid lines are fits of the measurement points to six Gaussian peak functions with a constant baseline intensity. Note the peak heights and widths are drastically altered between a) and b). All images are normalized to a fixed number of standard monitor counts to enable direct comparison between images, with the same scale for the color plots. Note the colorbar is a log-scale of intensity.}\end{figure*}

Alternatively, the compositional disorder inherent to the unit cell of the Co$_{x}$Zn$_{y}$Mn$_{20-x-y}$ series of materials may manifest as random local variations in the exchange field, which can serve as nucleating and pinning defect sites, irrespective of the crystallinity of the material. Examples include single crystal samples of doped (Fe, Co)Si, which have been shown to produce disordered chiral jammed states \cite{dustin}.  In particular, the internal chemical disorder and weak magnetocrystalline coupling produces multiple degenerate helical domains, which transform into skyrmion domains with weak relative orientation, upon application of a laboratory field \cite{dustin}. This leads to orientational disorder of the second length scale, wherein one domain will propagate, becoming trapped upon intersecting another domain and/or defect. This trapping precludes long-range order, precipitating intermediate chiral spin textures due to the domain spacing not satisfying integer skyrmion lattice constant multiples, resulting from randomly distributed nucleating sites \cite{dustin}. The inability of these structures to propagate or reorient leads to "jammed" configurations, such as chiral or labyrinth domains, amongst skyrmion domains \cite{dustin}. Therefore, the chemical disorder collectively generates a rich energy landscape exhibiting intermediate chiral and jammed skyrmion states.

Similarly, Co$_{x}$Zn$_{y}$Mn$_{20-x-y}$ possesses both internal chemical disorder \cite{metastable,spin} and relatively weak magnetocrystalline anisotropy around room temperature \cite{prebinger2021vital}. The ring-like scattering pattern displayed in Fig. 1c may therefore be attributed to the misalignment and trapping of the skyrmion and chiral domains which breaks the order in many directions. In reciprocal space, this would produce a ring scattering pattern due to the superimposition of rotationally offset hexagonal patterns and other magnetic structures, such as rotationally disordered helical domains, with no net orientation, but the same preferred q. Therefore, the degree of order may be interpreted from the angular width of the diffraction peaks, which we will examine later. Moreover, it is not to be ignored that we are in an above-room temperature phase; thermal fluctuations which stabilize this phase may also contribute fluctuation-disorder by way of skyrmion merging, splitting, collapse, and nucleation.

The skyrmion-ordering sequence which was used in the SANS experiments entails rotating the sample in the static magnetic field (about the y-axis in Fig. 1a), with the rotational range determining the efficacy of the skyrmion ordering response. This technique has been previously demonstrated to disentangle jammed chiral states, thought to reduce defect densities and nucleate additional skyrmion topological charge \cite{dustin}. This sequence was used in combination with various FC procedures and samplings of the phase space of the material (outlined in Fig. 2). Upon application of the ordering sequence in the disordered thermal equilibrium triangular lattice skyrmion phase ((D)E-T-SkX), the development of a first-order ring with 6 peaks was observed, accommodated by an additional secondary ring mimicking the same hexagonal symmetry with 12 peaks. Fig. 3a and b show the developing, and fully discernible 6-fold primary rings after 10 and 40 rotations, respectively. The presence of the secondary ring indicates potential multiple scattering and/or higher order diffraction. Comparing Fig. 1c and Fig. 3,  we observe the conversion of the disordered phase to the ordered triangular lattice phase via the dissolution of the ring diffraction pattern, and promotion of hexagonal peaks. Looking at the azimuthal projection of the primary diffraction rings of Fig. 3, we see that this transition manifests as an evolution of six peaks, accompanied by decreasing peak widths and baseline intensity. Note that there is no direct conversion between the decrease in ring intensity and increase in peak intensity, reinforcing the conclusion that the ring is comprised of additional chiral magnetic structures that upon rotation nucleate additional topological charge structures. This action is consistent with disentangling a jammed state of chiral domains, through the collective reorientation of skyrmion domains and formation of new skyrmions via the breakup of labyrinth domains through emergent monopole nucleation and propagation \cite{dustin,milde2013unwinding}. Therefore, the reported behaviour in Fig. 3 occurs as the rotational skyrmion alignment improves, and previously jammed states comprised of mixed helical/labyrinth phases are disentangled, enabling further skyrmion nucleation and propagation, which additionally contributes to the sixfold scattering signal.

\subsection{\label{sec:level2} Disordered metastable skyrmion transitions\protect\\}

As we cooled the sample by a FC procedure through the (D)E-T-SkX phase into the disordered metastable triangular phase ((D)M-T-SkX), there was a subtle development of peaks atop the ring diffraction pattern. The promotion of these peaks was observed to alternate with continued cooling, dissolving peaks into the background ring, while enhancing new peaks. The SANS measurement in the disordered square phase (Fig. 4) shows a heavily smeared four-fold pattern with significant promotion of the peaks along a preferred orientation of the q-vector. Subsequent re-warming processes into the room temperature skyrmion phase, in addition to AC susceptibility measurements performed in the square phase, confirmed the presence of skyrmions which contributed to the square pattern. The annular average confirms the presence of four peaks; the variable baseline intensity, height, and width of the peaks suggests preferential development of the square lattice phase along a preferred q-direction. That is to say, the two peaks along the anisotropy direction display decreased baseline intensities and peak widths, and increased scattering intensities compared to the remaining two peaks. This is consistent with recent studies examining cubic anisotropy in the Co$_{x}$Zn$_{y}$Mn$_{20-x-y}$ skyrmion series, which found that anisotropy controls the angular distribution of the q-vectors, exhibiting a trend with temperature similar to that of the inverse of the full width at half maximum of the SANS peaks \cite{prebinger2021vital}. In our case, however, the mere presence of four-fold peaks suggests a more complex action, generating improved rotational alignment in addition to increased development of the skyrmion tubes in the preferred q direction---whether it be through lengthening of existing tubes and/or nucleation of new tubes, contrary to what one would expect. In \cite{prebinger2021vital}, increased anisotropy enhances orientational order and defines the distribution in rotational alignment of the  skyrmions. Alternatively, for the case of disordered skyrmions there is this additional mechanism of skyrmion development, wherein anisotropy may play a greater role by facilitating the disentangling of jammed states to allow reorientations to a square phase. We note that the overall scattering intensity of the square lattice peaks are greatly reduced from the ordered triangular lattice, a fact which may be attributed to disorder somewhat inhibiting skyrmion development and reorientations, resulting in leftover topologically trivial states such as helical, conical, and ferromagnetic domains. Neither conical or ferromagnetic states would contribute to the scattering signal as the ferromagnetic signal is masked in the central peak by the beam block, while conical domains would have to be probed in a perpendicular field geometry.

\begin{figure}
\includegraphics[width=\columnwidth]{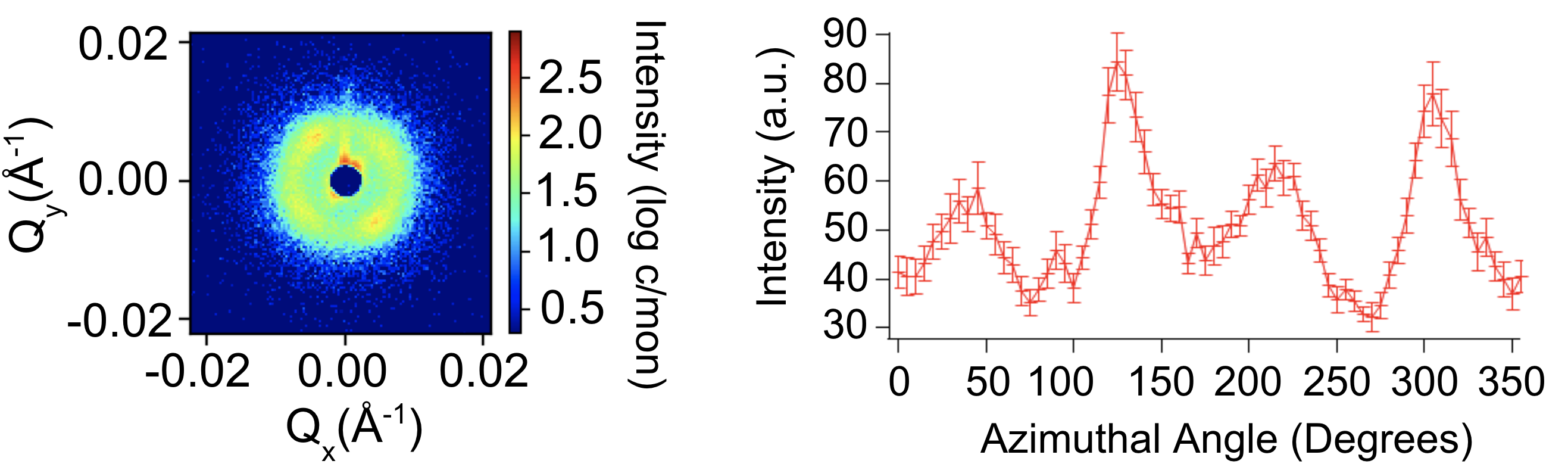}
\caption{ SANS image, and it's corresponding annular average, for the square lattice skyrmion phase, with significant promotion of the peaks along the anisotropy direction. Solid lines are fits of the measurement points to four Gaussian peak functions with a constant baseline intensity. Measurement was taken at 28 K in a field of 100 G. Note the colorbar is a log-scale of intensity. }\end{figure}

 A decrease in the A/D ratio is stipulated to accompany the increase in anisotropy upon decreasing temperature. These two scalings are thought to trigger the triangular to square lattice transition through skyrmion elongation by directional expansion and subsequent rearrangement \cite{robust,skyrmion_frac}. The interplay of topological stability with an enhancement of q demands spin textures which satisfy conservation of skyrmion number between triangular and square phases. In thin plate samples, this is accomplished via the directional expansion of skyrmions into deformed bar and L-shaped elongated textures whose directionality is determined by magnetic anisotropy \cite{deformationMorikawa2017,elementukleev2019}. In bulk samples, conservation of skyrmion number can be achieved through solutions which increase the volume occupied by each skyrmion, such as in deformed textures similar to \cite{deformationMorikawa2017,elementukleev2019}, or alternatively through solutions which decrease the volume occupied by the square lattice of skyrmions.

 For our sample, the ratio of skyrmion density in the triangular phase at 310 K, to the square phase at 28 K, is given by $\frac{\rho_{S}}{\rho_{T}}=\frac{\sqrt{3}}{2}(\frac{Q_{S}}{Q_{T}})^{2}$ \cite{metastable}, where ${\rho_{S}}$ and ${\rho_{T}}$ are the skyrmion densities in the  square and triangular phases, and ${Q_{S}}$ and ${Q_{T}}$ are the positions of the magnetic reflections for the square and triangular lattice forms in q-space. This expression derives from considering the ratio in packing density for an ideal triangular lattice ($\frac{\pi}{2\sqrt{3}}a_{T}^{2}$) versus square lattice ($\frac{\pi}{4}a_{S}^{2}$) given their lattice periodicity's, $a_{T}$ and $a_{S}$, as defined by $Q_{T}$ and $Q_{S}$ in q-space. The q-centers for each phase were extracted as the peak centers from Gaussian functions fit to radial averages of the SANS images. This yields a ratio of $\frac{\rho_{S}}{\rho_{T}}\approx$ 1.78, consistent with \cite{metastable}. However, to keep skyrmion density constant between idealized triangular and square lattice phases, the square lattice constant should decrease by a factor of $\sqrt{\frac{2}{\sqrt{3}}}\approx 1.075$ corresponding to a  square q value of 0.0056 \AA $^{-1}$, contrary to our observed value of 0.008 \AA $^{-1}$. Thus the skyrmion density violates conservation of topological charge when assuming a perfectly ordered square lattice. This can be resolved through solutions which decrease the volume occupied by the square lattice by incorporating additional magnetic structures with no topological charge that are defined by the same orthogonal double-q vectors, such as with helical or conical domains.
 
 In the special case of a disordered square lattice transition pathway, the emergence of some net four-fold order when coming from a jammed labyrinth state suggests three possible physical cases. In the first case, skyrmions are nucleated from the mixed helical state present in the disordered phase, with their alignment determined by the direction of increased anisotropy. In the second case, an in-plane elongation and subsequent reorientation of the tubes may partially disentangle trapped domains and jammed states along distinguished direction, nucleating new skyrmions. In the final case, the in-plane elongation of the jammed state through merging and subsequent reorientation yields a deformed nematic-like square texture, similar to \cite{elementukleev2019,deformationMorikawa2017}. In all of these cases it appears that, in order to overcome some of the jamming inherent to the disordered state, a change in topology is required in order to allow for skyrmion restructuring/reorientations through elongation, merging, and nucleation processes. For our material, we stipulate that mechanism (2) or (3) produces skyrmion reorientations that disentangle jammed states---improving skyrmion alignment and possibly nucleating oriented skyrmions upon disentangling. For mechanism (1), the magnetocrystalline anisotropy would have to increase to a large enough value so as to stabilize new skyrmions well below $T_{c}$ however, for this doping series of materials, the only disconnected skyrmion phase observed at low temperatures is a disordered phase stabilized by magnetic frustration \cite{disordered}. Whereas in support of (2) and (3), we have already observed skyrmion reorientations associated with anisotropic considerations \cite{dustin,henderson2020characterization}. Note that mechanism (2) could still produce skyrmion textures reminiscent of (3) when transitioning in a disordered lattice, as skyrmions reorient and intersect jammed helical/skyrmion states, resulting in an elongated deformed skyrmion texture along a preferred q-direction consistent with anisotropy.

 \begin{figure}
\includegraphics[width=\columnwidth]{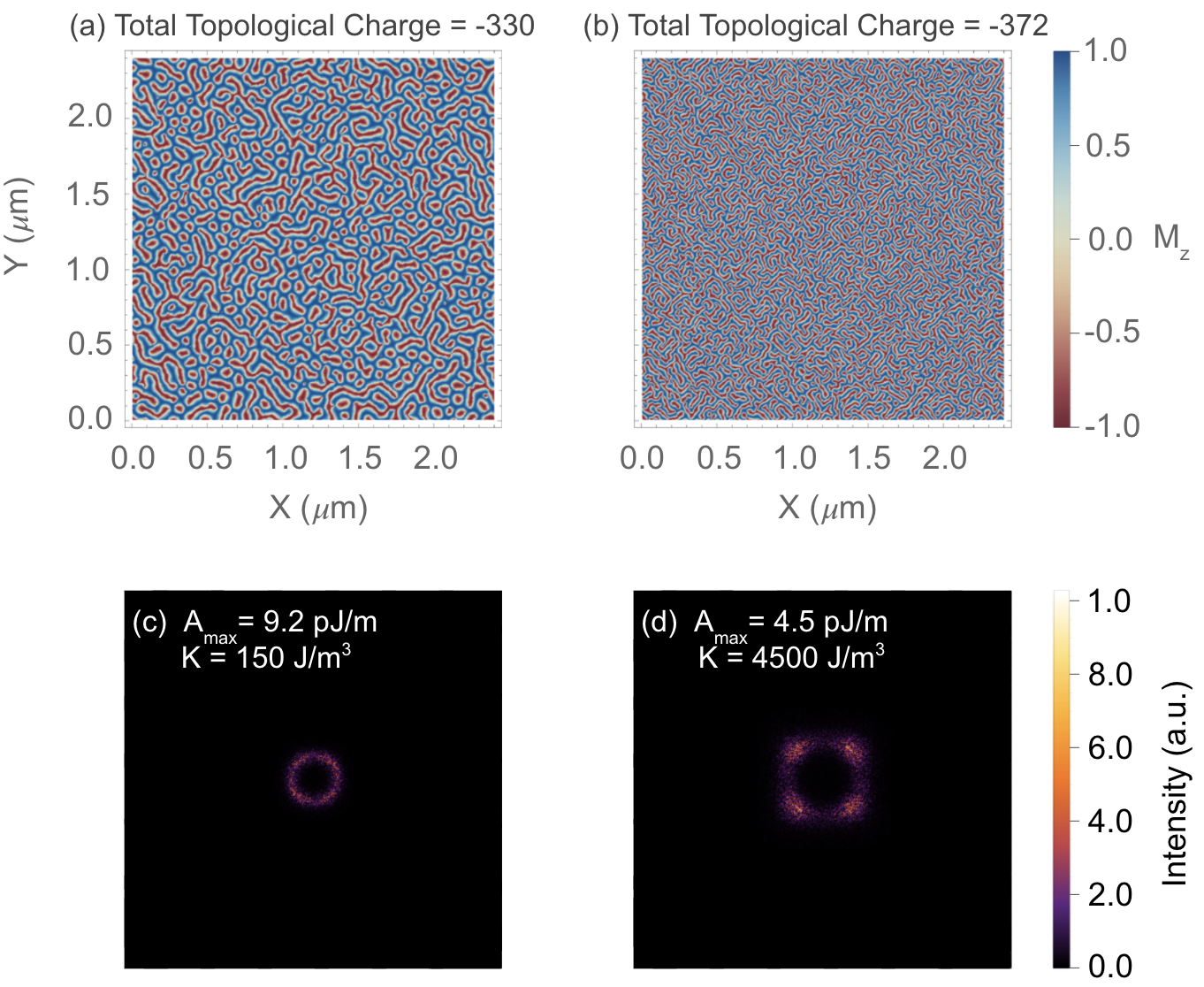}
\caption{ Micromagnetic simulations and their corresponding SANS patterns for a thin-plate Co$_{8}$Zn$_{8}$Mn$_{4}$ sample of dimensions 2.4 x 2.4 $\mu m^{2}$. Out-of-plane magnetization, $M_{z}$, is shown for the simulated triangular-to-square lattice transition starting from the initially disordered phase (a) phase at 310 K, to the disordered square phase at 30 K (c). The total topological charge per 2D slice is observed to increase during the transition by approximately 12\%. Corresponding SANS images demonstrate an increasing scattering ring size, consistent with a reduction in $A_{max}$, while the development of four-fold order is observed in plot (d).} \end{figure}

Micromagnetic simulations performed on a disordered state suggest a transition pathway which entails a change in topology, contrary to observations made for ordered lattices \cite{deformationMorikawa2017,elementukleev2019}. Simulations were performed using the Ubermag micromagnetic simulation package \cite{beg2022} on a lattice of size 2400 nm x 2400 nm x 5 nm, with a discretization cell size of 5 nm. A random spin configuration was initialized in an external magnetic field of 400 mT (along the z-direction) to mimic the experimental field-cooling procedures. A DMI constant of 0.00053 J/m was used \cite{spin-wave,elementukleev2019}. To simulate a disordered state with site-by-site disorder, a varying scalar exchange field was used, set to vary between 0 and a maximum value $A_{max}$. The equilibrium-square metastable lattice transition was performed by incrementally decreasing $A_{max}$ from 9.2 pJ/m to 4.5 pJ/m to simulate the relative decrease in A/D ratio, while a cubic anisotropy term was simultaneously increased from 150 J/$m^{3}$ to 4500 J/$m^{3}$. The former change induces an enhancement of the q-vector, while the latter favors a particular elongation direction. While the decrease in A/D ratio may manifest a change in the helical periodicity similar to experiment, the primary origin of the change in q-vector is thought to be rooted in the low-temperature antiferromagnetic frustration of the Mn sublattice which can shorten the helical pitch and deform skyrmions \cite{elementukleev2019,SpinUkleev2022}. Similar competition between DMI and frustrated exchange interactions have been shown to produce short-size magnetic modulations \cite{SkyrmionMutter2019,Fujishiro2019}. These simulation parameters were taken from studies which determined exchange and cubic anisotropy values as a function of temperature through soft x-ray scattering measurements coupled with simulations, and ferromagnetic resonance techniques \cite{elementukleev2019,prebinger2021vital}. Relaxation times of 0.5 ns were introduced between each step. A two-dimensional magnetization slice of the out-of-plane magnetization $M_{z}$ for the initial disordered state shows a mixed skyrmion and helical/labyrinth state, with a topological charge of -330 (Fig. 5a). A corresponding simulated SANS image exhibits a roughly uniform ring-like scattering pattern (Fig. 5c).  Cooperatively decreasing $A_{max}$ while increasing K shows results in a q-vector magnitude dependence consistent with our SANS data and $4\pi\frac{A_{max}}{D}$ which defines the helical pitch. Magnetization slices show a gradual merging and separation of labyrinth states, producing deformed skyrmions elongated along two orthogonal double-q vectors. These restructuring dynamics are in direct opposition to LTEM and micromagnetic simulation studies performed on ordered skyrmion states in \cite{deformationMorikawa2017} and \cite{elementukleev2019}, respectively. In the former two studies, deformation and elongation of the skyrmions along magnetic easy axis obeys conservation of topological charge. The deformations ensure the occupied skyrmion volume is constant despite the changing lattice shape and periodicity, thereby conserving skyrmion density. Conversely, our simulations demonstrate an increase in topological charge from -330 to -372 during the square lattice structural transition as skyrmion and labyrinth structures merge and dissolve, allowing skyrmion elongations and reorientations along preferred directions, which ultimately produce deformed textures reminiscent of \cite{deformationMorikawa2017}. The simulated and experimental SANS patterns exhibit a similar conversion from a ring-like to four-fold pattern. 
These results reinforce the requirement of a change in topology when undergoing disordered-to-ordered square structural lattice transitions. This observation suggests the disordered square lattice transition takes place by a unique pathway, not accessible by ordered states, which involves the merging and separation of helical and labyrinth states. This restructuring enables skyrmion nucleation, elongation, and  reorientations necessary for disordered square lattice structural transition.

\subsection{\label{sec:level3} Ordered metastable skyrmion transitions \protect\\}

For the ordered thermal equilibrium triangular lattice skyrmion phase ((O)E-T-SkX) shown in Fig. 6a, transitioning to the lower temperature phase marked the development of higher order scattering rings up to the third order, while the baseline ring intensity approached zero (Fig. 6b). The significant reduction in peak widths and baseline intensity in Fig. 6a versus Fig. 3b points towards improved rotational alignment and development of the skyrmion lattice. We attribute this to the increased rotational range of the ordering sequence, consistent with \cite{dustin}. Similarly, the increase in scattering order and decrease in baseline intensity between the thermal equilibrium (Fig. 6a) and corresponding ordered metastable triangular lattice skyrmion phase ((O)M-T-SkX) (Fig. 6b) implies a lengthening and/or nucleation of the tubes and therefore increase in correlation lengths, possibly due to reduced thermal fluctuations and increased pinning/anisotropy contributions. The presence of additional scattering rings are most likely attributable to a combination of multiple scattering and higher-order scattering processes. 

\begin{figure}
\includegraphics[width=\columnwidth]{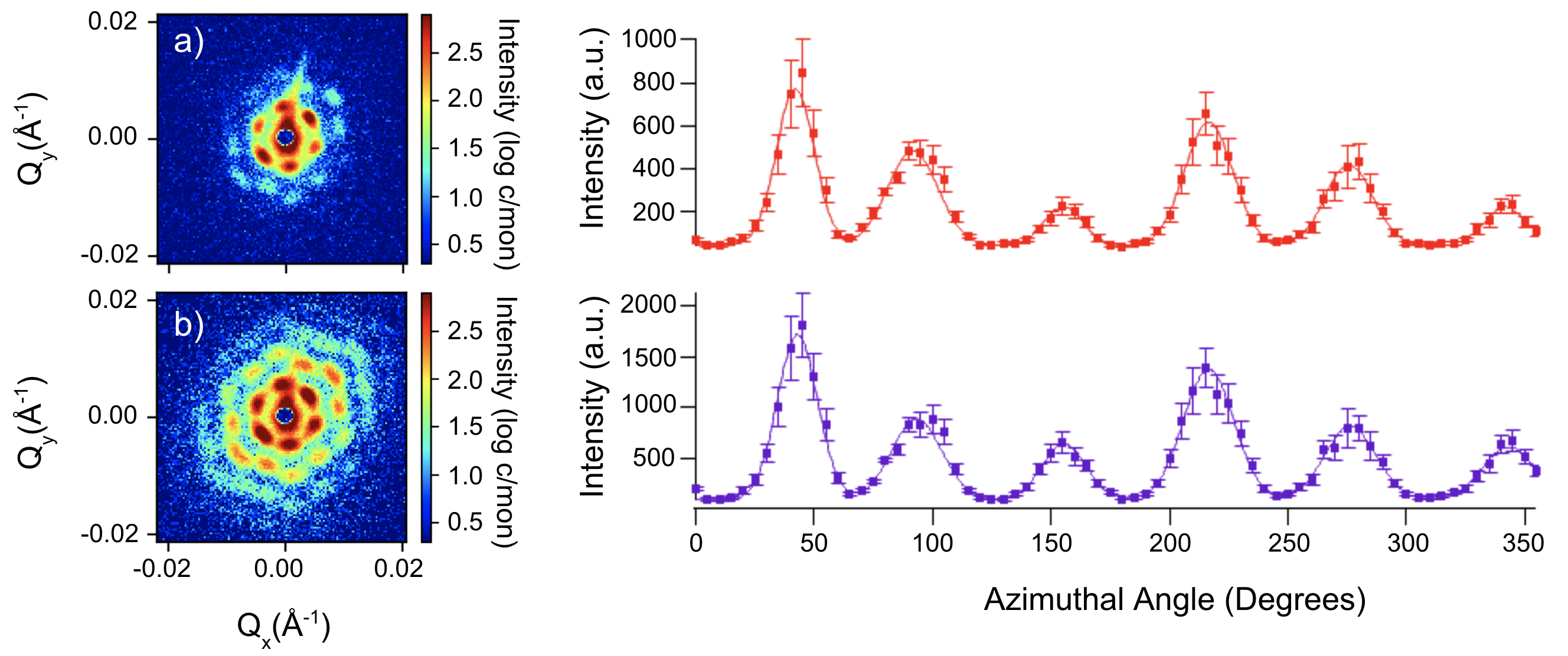}
\caption{ SANS diffraction patterns and their corresponding annular averages of their primary scattering rings for a) the sample after 20 rotations of the ordering sequence at 310 K in a magnetic field of 250 Oe, and b) after cooling to 173 K and 122 G from a). Solid lines are fits of the measurement points to six Gaussian peak functions with a constant baseline intensity. Note the baseline intensity essentially goes to zero from a) to b), and we observe the development of higher order scattering rings the more we cool into the (O)M-T-SkX phase. Both images are normalized to a fixed number of standard monitor counts to enable direct comparison between images, with the same scale for the color plots. Note the colorbar is a log-scale of intensity.}\end{figure}

 \begin{figure*}
\includegraphics[width=\textwidth]{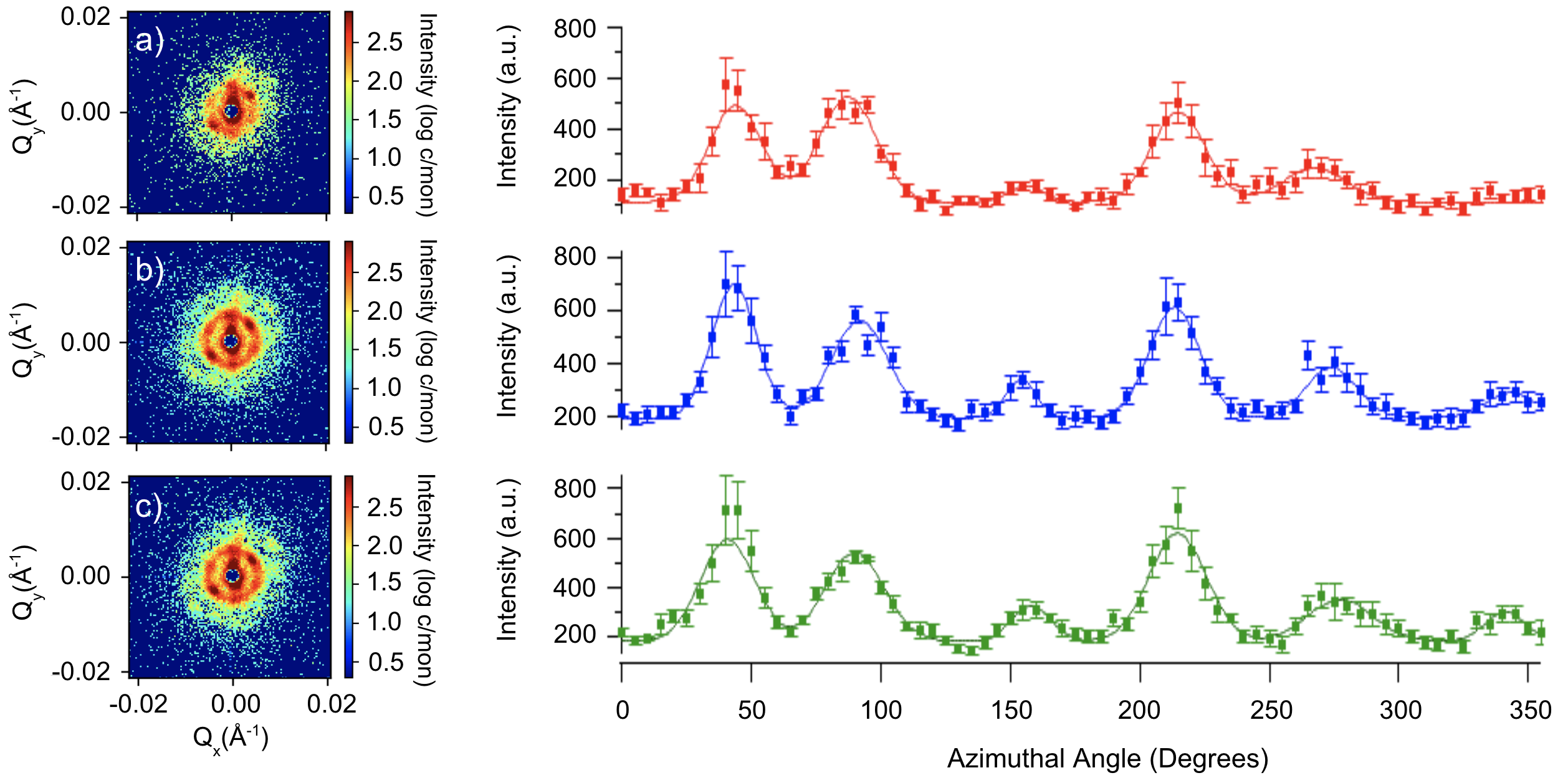}
\caption{ SANS images, and their corresponding annular averages of their primary diffraction rings for the (O)M-TSkX phase at 173 K upon a) lowering the field back to 10G after 
saturation into the ferromagnetic phase in a field of 1T, b) after 10 ordering sequence rotations in a field of 10 G, and c) after 10 ordering sequence rotations in a field of 250 G. Solid lines are fits of the measurement points to six Gaussian peak functions with a constant baseline intensity. All images are normalized to a fixed number of standard monitor counts to enable direct comparison between images, with the same scale for the color plots. Note the colorbar is a log-scale of intensity.  }\end{figure*}

Remarkably, saturation of the metastable skyrmion lattice into the ferromagnetic phase generates a memory effect upon re-entry into the metastable skyrmion field window. Namely, a diffraction pattern reminiscent of the pattern prior to saturation is observed after lowering the field, with similar azimuthal peak positions and a prominent anisotropy direction (Fig. 7a). This reemergence of a skyrmion diffraction pattern in the metastable phase of a previously annihilated skyrmion lattice suggests that a memory of the skyrmion lattice is present, even in fields high enough to destroy all skyrmion and spiral scattering signals. This memory effect is quite surprising given the underlying disorder of the material which tends to precipitate jammed, rotationally disordered states. The competition between this disorder and the tendency of skyrmions to nucleate and maintain their previous orientations is apparent in the slightly broadened peak widths. Previous FC procedures from the (D)E-T-SkX through the metastable phase demonstrated a disordered ring SANS pattern for the same field and temperature. The skyrmion order parameter provides an additional measure of the memory effect. In particular, not only does the state after saturation have to produce topological skyrmion states, but it must overcome the natural energetics of the phase which favors disordered states. Relaxation into a disordered state after saturation would be possible if there was a slight memory effect set by low-temperature pinning sites, with jamming energetics still dominating. However, the reproduction of previous skyrmion order, which was generated in a separate phase, overcomes the naturally disordered ground state of the metastable skyrmion phase. The generation of this ordered state therefore requires additional energetics not available in the metastable phase. Therefore, the degree of order that is retained after saturation, in spite of the phase's underlying tendency towards disordered ring-like states, strongly suggests a memory effect.

Examining the influence of the ordering sequence on the memory effect provides an indication of the energetics of the phase. The ordering sequence was applied in fields of both 10 G (Fig. 7b) and 250 G (Fig. 7c). The baseline intensity is observed to slightly increase after the first ordering sequence, while the third and sixth peaks are further enhanced, reproducing previous peak asymmetries. This suggests the sequence may in fact orient skyrmions and helices, while also nucleating additional skyrmions from their pinning sites. These skyrmions appear to be both ordered and disordered, contributing to the peak and baseline intensities, respectively. This illustrates the dual nature of disorder: it may both enhance skyrmion development through defect pinning sites while also impeding reorientations and alignment through jammed chiral textures. Interestingly, there is significant development of the second order rings between ordering procedures, possibly indicating a lengthening of the skyrmion tubes, thereby reinforcing \cite{toron}.  It is possible that the varying interplay of the magnetic field direction and anisotropy direction encourages an anisotropy favored elongation of magnetic torons into skyrmions \cite{toron}. Torons may be visualized as skyrmion fragments that terminate their prolongation on bloch points \cite{toron}. These torons may exist as remnants of the previous skyrmion lattice \cite{memory}. Their survival is likely enhanced by pinning on defects. These defects may be naturally present in the material owing to internal disorder, which may be rooted in the site mixing of atoms \cite{birch2020real} or presence of grain boundaries \cite{MnSigrain}.

\begin{figure}
\includegraphics[width=\columnwidth]{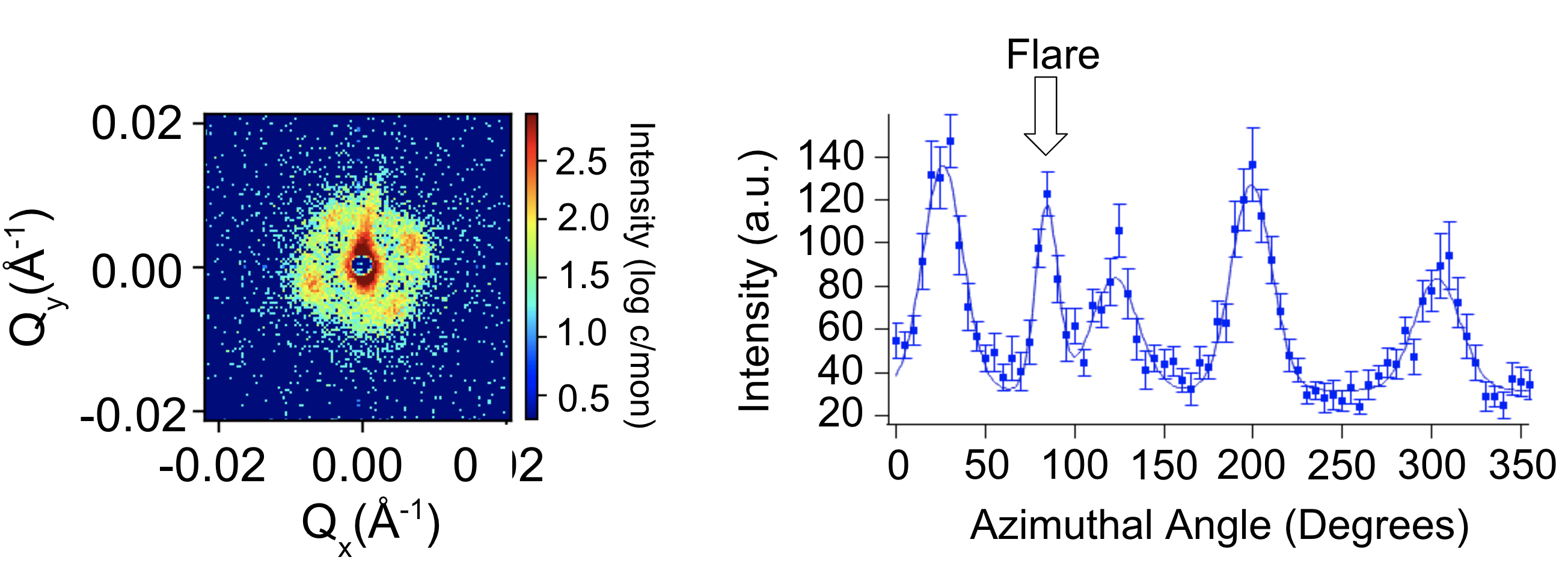}
\caption{ SANS image and it's corresponding annular average, for the ordered square lattice skyrmion phase. Solid lines are fits of the measurement points to five Gaussian peak functions with a constant baseline intensity. Measurement was taken at 60 K in a field of 6 G. The annular averages contains 5 peaks; the additional peak is due to a scattering flare from the surface of the sample. Note the colorbar is a log-scale of intensity. }\end{figure}

\begin{figure*}
\includegraphics[width=\textwidth]{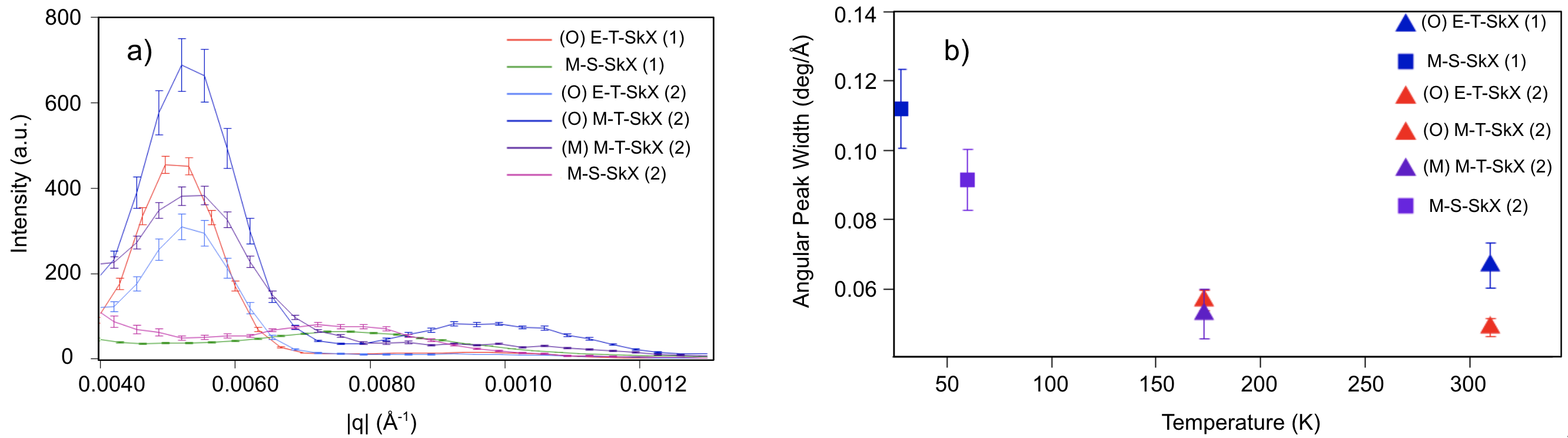}
\caption{ Radial $|q|$ dependence of SANS intensity, integrated azimuthally over entire SANS image for q $\geq$ 0.0040 \AA$^{-1}$ (smaller q was excluded due to the presence of flares) for triangular, metastable, memory, and square skyrmion phases (a). Angular peak width as a function of temperature for triangular, metastable, memory, and square skyrmion phases (b). Angular peak width is defined as the q value of the primary diffraction rings as determined from radial averages a), multiplied by the average peak widths extracted from the annular averages in Figures 4-8.   }\end{figure*}

 Upon further cooling, the square phase was entered, yielding four discrete peaks (Fig. 8). Comparing Fig. 8 and Fig. 4, we observe significantly increased scattering and discretization of the peaks for the square phase entered through the ordered phase, versus the disordered phase, emphasizing the role of previous order and/or pinning in structural lattice transitions. Most likely, the cooperative increase in anisotropy and decrease in A/D drives a collective reorienting and elongation of skyrmion tubes, leading to an increase in scattering along distinguished directions, consistent with \cite{deformationMorikawa2017,spin-wave}. The history dependence of the structural lattice transition on previous skyrmion order supports separate lattice transition pathways for ordered versus disordered skyrmion states. Namely, internal disorder has a strong influence on a skyrmion states ability to reorient, invoking different restructuring dynamics which may change the topology of the system.

Figure 9a shows the azimuthally integrated radial $|q|$ dependence for triangular, metastable triangular, memory, and square skyrmion SANS images. All forms of triangular skyrmion lattice, i.e. equilibrium, metastable, and memory, display similar peak centers. Likewise, the two square lattice phases also exhibit similar peak centers, shifted to larger q relative to the triangular ones, consistent with \cite{robust,metastable}. The two square phases display comparable scattering intensities, while the triangular phases are more varied in their respective scattering intensities. This larger variability in scattering intensity amongst the triangular lattice phases may be representative of the overarching diversity of stabilization mechanisms for the triangular versus square skyrmion phases, enabling the triangular lattice forms to achieve larger ranges of skyrmion development, i.e., formation, penetration, and alignment. Figure 9b demonstrates the degrees of order for the various skyrmion phases, encapsulated in the angular peak widths as a function of temperature. Angular peak widths were calculated using the average of the azimuthal peak widths of the primary scattering ring, multiplied by the radial q location of the peaks taken from fits to data in Fig. 9a. We note that although the two square phases are stabilized through two different pathways (i.e., ordered versus disordered), and display varying degrees of order, their comparable intensities may suggest square lattice formation occurs via a similar driving mechanism for this particular compound, with limited development/proliferation throughout the material. It is indeed possible that the square phase is not the majority phase, and is instead stabilized at surfaces, grain boundaries, and/or defects, resulting in its bounded formation irrespective of its transition pathway. If both the ordered and disordered square phases exhibit similar final states, the lattice transition must produce comparable skyrmion elongation and reorientation in the ordered and disordered states, precipitating similar volume fractions. This may suggest the presence of a dominant conical domain in the square lattice phase. There also appears to be an increase in diffuse scattering in the disordered square phase, suggestive of increased local disorder. This may result from disorder-enhanced q-vector fluctuations brought on by the deformation of the disordered triangular lattice state during the square lattice transition. Conversely, the presence of diffuse scattering in the thermal equilibrium phase may be indicative of thermal fluctuations or coexisting precursor helical or conical phases near the skyrmion phase boundary.

Estimates of the relative skyrmion volume fractions for the (O)M-T-SkX vs (M)M-T-SkX phases can be generated by comparing the summed radial $|q|$ intensities. For a magnetic skyrmion sample, the scattered neutron intensity will depend on the saturated magnetization, the correlation length of the magnetic structures, and the volume fraction occupied by those structures. As an approximation, we may perform SANS simulations on magnetization arrays of various skyrmion volume fractions, with a fixed correlation length, to examine their independent influence on scattered neutron intensity. Magnetization arrays of size 640 nm x 640 nm x 2048 nm were generated with discretization cell sizes of 5 nm using the micromagnetic simulation package Ubermag \cite{beg2022}. Assuming contributions to the hexagonal diffraction patterns arise solely from skyrmionic structures which occupy the entire lattice volume, we may take their ratio in scattered intensities and use this to determine the approximate difference in skyrmion volume fraction. We created composite magnetization arrays of skyrmion domains with a ferromagnetic domain of varying volume fraction. We concern ourselves primarily with coexisting skyrmion and ferromagnetic domains, since we transitioned directly from the ferromagnetic to skyrmion state without crossing the zero field boundary. In doing so, we can estimate the relative change in skyrmion volume fractions for the ordered metastable skyrmion phase versus its memory counterpart. Hysteresis and low-temperature trapping of ferromagnetic domains is possible after saturation, with ferromagnetic contributions to the scattered intensity  masked by the beam block, resulting in a reduced scattered intensity. While other chiral and/or non-chiral domains such as conical and helical are possible, decoupling the multi-variable contributions of the domain types and their correlation lengths is beyond the scope of this paper. Conical domains could similarly appear to reduce the scattered neutron intensity, with contributions only accessible in the perpendicular field geometry. Alternatively, helical domains could both increase and decrease the SANS intensity as a complex function of their longitudinal correlation lengths.

\begin{figure*}\center
\includegraphics[width=\textwidth]{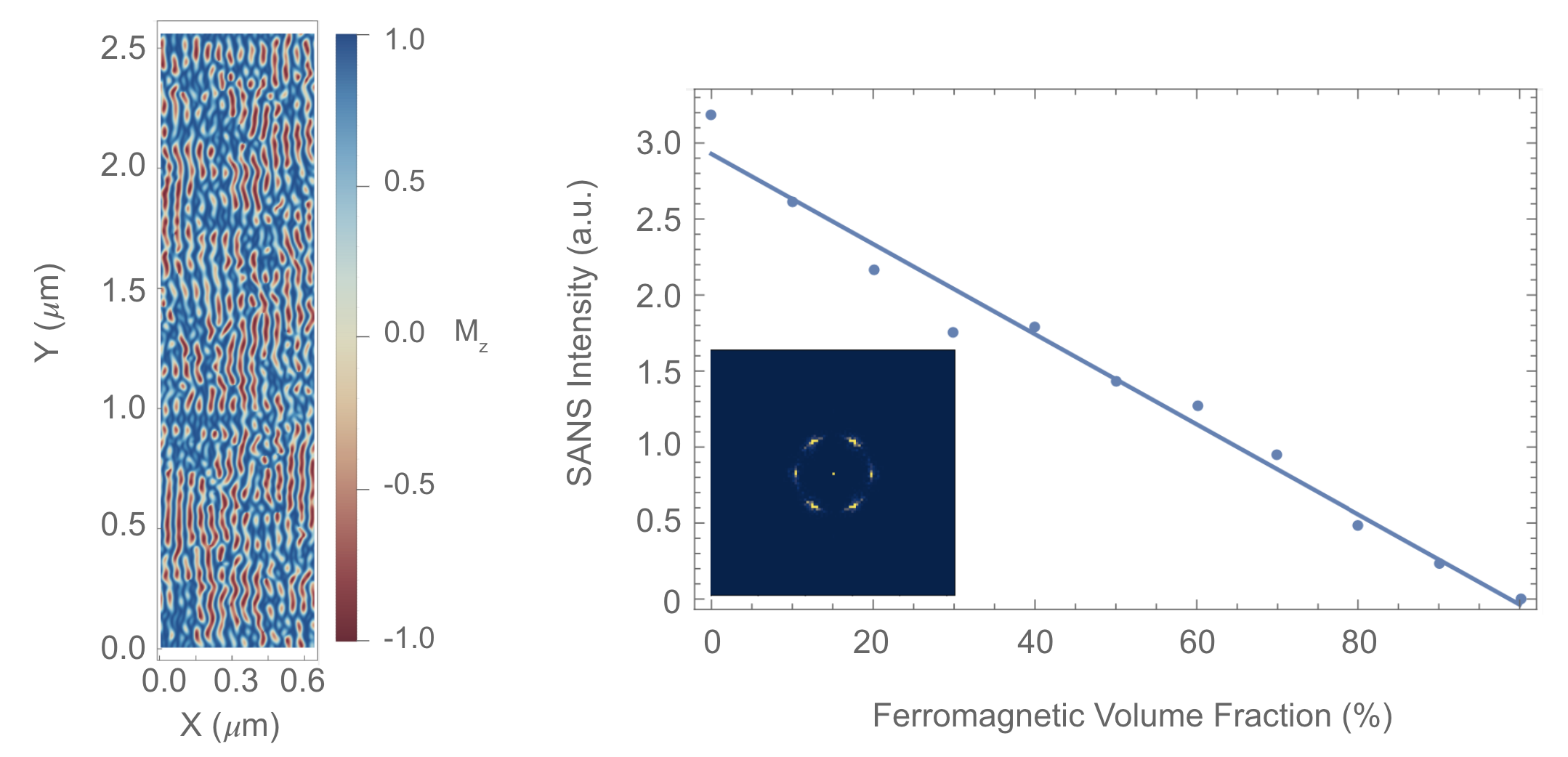}
\caption{  Simulations demonstrating the relationship between skyrmion to ferromagnetic volume fraction and scattered neutron intensity. XZ magnetization slices of the out-of-plane magnetization component, $M_{z}$, are displayed for skyrmion (left) and ferromagnetic (right) magnetization volumes. The simulated SANS intensity is plotted as a function of the ferromagnetic volume fraction in the magnetization array. A simulated SANS image is displayed in the bottom left corner of the plot for a composite magnetization arrays consisting of a triangular skyrmion lattice and ferromagnetic domain with 100 \%  and 0\% volume fractions, respectively. The SANS intensity demonstrates a linear relationships which decreases for increasing ferromagnetic volume fractions.}\end{figure*}

SANS simulations were performed for fixed size polydomain magnetization volumes with triangular lattice skyrmion volumes and ferromagnetic volumes shown in Fig. 10. The scattering intensity was examined for varying skyrmion lattice and ferromagnetic volume fractions, integrating the scattering intensity over the first order scattering ring. The scattering intensity was observed to decrease for increasing ferromagnetic volume fractions, displaying a linear relation (Fig. 10). The experimental SANS images were integrated over an annular region from q=0.0040 \AA$^{-1}$ to q=0.0080 \AA$^{-1}$ capturing the entire first order scattering ring, while excluding contributions from flares and higher orders. The ratio of the intensity in the (O)M-T-SkX (2) to (M)M-T-SkX (2) phases is $\simeq$ 1.4.  Assuming the (O)M-T-SkX phase has a skyrmion volume fraction of 95 \%, we can estimate a decrease in the skyrmion volume to 68 \% in the memory state.

\section{\label{sec:level3}Discussion\protect\\}

Our SANS measurements demonstrate three distinct regimes in which thermal fluctuations, pinning, and anisotropy terms dominate skyrmion ordering responses and reorientations. A collective reorientation takes place in the thermal equilibrium phase, while pinning and memory effects are observed to enhance skyrmion stability in the metastable triangular phase. A new disordered-to-ordered skyrmion square lattice structural transition is revealed through elongations which necessitate a change in topology in order to enable reorientations of the jammed state to a deformed square pattern. This is reinforced through micromagnetic simulations on a disordered lattice as exchange and anisotropy parameters are incrementally varied, demonstrating an increase in topological charge during the transition through the breakup of labyrinth domains which nucleates deformed skyrmions. Together, these observations emphasize the fundamental mechanisms and interplay of magnetic anisotropy and defects in skyrmion stabilization and structural lattice transitions.

One might expect a disordered triangular skyrmion lattice to display similar degrees of disorder in the square lattice phase (b.3 of Fig. 11). In the extreme case, such disorder may even inhibit the triangular-to-square structural lattice transition, where pinning defects and trapped chiral domains preclude skyrmion elongation and reorientations, and suppress the monopole motion required for the breakup or merging of labyrinth domains. On the other hand, a finite amount of disorder may facilitate skyrmion ordering and transitions by lowering the barrier for monopole-antimonopole creation. In addition to the influence of disorder on square lattice transitions, the cruciality of anisotropy in such transitions has been reinforced across multiple studies. In MnSi the triangular-to-square lattice transition in quenched skyrmions entails a pathway in which the combination of reduced net magnetization upon decreasing magnetic field, and favored magnetic moment directions due to anisotropic perturbations, produced skyrmion reorientations which generate the square lattice \cite{MnSi_transition}. Additional triangular-to-square lattice pathways invoke an increase in easy-plane anisotropy, in which anisotropy favored growth of skyrmions leads to skyrmion overlap, and the subsequent triangular-to-square lattice transition \cite{skyrmion_frac}. Most recent studies for this material suggest that the role of anisotropy in ordered square lattice transitions is to set the preferred direction of distortion of the peaks.  However, our simulations reveal that the combination of varying exchange and anisotropy is required for the new disordered square transition pathway, which invokes a change in topology. As mentioned previously, anisotropy may contribute to elongation and reorientations, while disorder-related defects may lower the energy required for topological transitions, overcoming the topological protection of skyrmions.

The memory of the previous skyrmion lattice in the (O)M-T-SkX phase persisted in spite of saturation to the ferromagnetic phase---marked by the disappearance of all skyrmionic and spiral scattering signals. The recovery of the same approximate azimuthal peak positions and relative peak intensities, despite the predisposition of the skyrmion phase to disordered and jammed states, further underscores the origin of this phenomena to be memory based. One might expect such a memory effect, owing to the ``frozen in" nature of the metastable phase, as shown in the upper plot c.3 of Fig. 11. One possible explanation for this apparent memory may lie in low temperature pinning phenomena and enhanced magnetic anisotropy, highlighting the fundamental stabilization energetics of skyrmions in the metastable phase. This result is consistent with skyrmion memory observations made in Cu$_{2}$OSeO$_{3}$, in which the memory of skyrmion lattice states persisted after increasing the magnetic field to the field polarized state \cite{memory}. This memory effect however, did not persist in zero field and was demonstrated in a well ordered triangular phase that did not require any previous ordering procedures.  Our memory effect is made substantially more robust by the fact that it exists in the absence of a magnetic field, and occurs in an inherently disordered material, demonstrating a complex competition of stabilization energetics. The latter may be highlighted by the slightly enlarged peak widths for the memory state. One may deduce the competition between previous pinning, which favors the same peak locations and widths, and underlying disorder which favors smeared ring-like patterns. The intensities of our peaks are, however, significantly reduced indicating underdeveloped skyrmion lattice formation as compared to it's original state. Simulations in which the skyrmion-to-ferromagnetic volume fraction ratio was varied suggest a decrease in the skyrmion volume fraction from an assumed  value of 95\% to 68\%. The remaining skyrmion volume fraction in the (M)M-T-SkX phase should be correlated with the disorder in the material, which determines the number of defect pinning sites, and therefore possible locations for toron survival. This suggests the polycrystallinity of the sample may contribute to the observed memory through grain boundary disorder and domain intersection. Examining the same memory effect for an ordered lattice would help elucidate the interplay of skyrmion memory and disorder (such as pinning defects, polycrystallinity, and complex jammed energy landscapes).

Skyrmion lattice correlations extending beyond the skyrmion envelope have also been shown to exist upon FC in   $\mathrm{Fe}_{1-x}\mathrm{Co}_{x}\mathrm{Si}$, producing a metastable phase, where intensities almost two orders of magnitude smaller than in the A phase are thought to indicate origins rooted in surface or edge pinning \cite{A_phase_memory}. On the other hand, previous studies using Magnetic Force Microscopy (MFM) on a polished surface of $\mathrm{Fe}_{1-x}\mathrm{Co}_{x}\mathrm{Si}$ have demonstrated the preferential and reproducible decoration of certain positions on the surface with metastable skyrmions during field-cooling runs, owing  to enlarged local potential barrier at defect pinning sites \cite{surface_pinning}. Given the chemical disorder inherent to this material, which stabilizes a high density of defects, it is possible that the skyrmion memory is encoded and survives in the form of isolated skyrmions (or torons as suggested in \cite{memory,FieldLeonov2020}) confined to preferential positions, attributed to defect-related pinning. One fashion in which the skyrmion lattice may then propagate between these isolated skyrmions, which set the lattice orientation, is by a monomer-by-monomer addition  type mechanism previously observed in \cite{MA}.  Theoretical evidence of isolated skyrmions have been shown to exist as metastable objects within the saturated phase for a broad range of uniaxial anisotropy in the phase diagram of \cite{Uni_anisotropy}. It would therefore follow that the underlying disorder in the material, both compositional and crystalline, plays a considerable role in the observed memory. Thus, the pinning of the skyrmions may be enhanced at lower temperatures and with increased anisotropy.

Studying composition effects via chemical substitution enables a pathway in which one can manipulate the delicate balance of stabilization energy terms, while introducing additional disorder and pinning effects. In particular, exploring the memory effect as a function of x,y concentration in Co$_{x}$Zn$_{y}$Mn$_{20-x-y}$ would illuminate the role of defects and anisotropy in the skyrmion memory phenomena. Additionally, one could examine the influence of disorder induced by grain boundaries in polycrystalline samples by performing the memory procedure for varying levels of sample crystallinity, with powdered samples used in the extreme case. Studying the persistence of the memory effect as a function of the strength of the saturating field could provide an estimate on the lower bound of the energy barrier of defect-related pinning. Moreover, exploring the memory effect as a function of the temperature and field of the metastable lattice, before saturation, would highlight the fundamental stabilization energetics of skyrmions in the metastable phase. Future experiments may also vary the time over which the magnetic field is reduced back to its skyrmion envelope value, and in so doing probe a characteristic time scale over which the skyrmion memory persists, similar to metastable skyrmion lifetime estimates in \cite{doping}. Noting that the (O)E-T-SkX phase does not exhibit the same memory effect upon saturation into the ferromagnetic phase---the disordered ring SANS pattern is regained---reinforces the disparate nature of the stabilization and formation mechanisms of the thermal equilibrium versus metastable phases; that is, thermal agitation, disorder, and topological stability. Performing simulations which examine lattice transition pathways and memory effects as a function of disorder levels would help establish a boundary between regimes of disorder which facilitate and inhibit skyrmion lattice transitions and stability. This could be used to establish ideal defect densities for enabling skyrmion reorientations and enhancing stability, guiding the tailoring of future material parameters for spintronic applications. We intend to further explore skyrmion formation and stabilization mechanisms across multiple phases in the bulk using a newly developed reconstruction algorithm \cite{heacock2018sub}, as well as incorporate spin dynamics \cite{nsofini2016spin,Sarenac2019,Sarenac2018}.

\begin{figure*}\center   
\includegraphics[width = \textwidth]{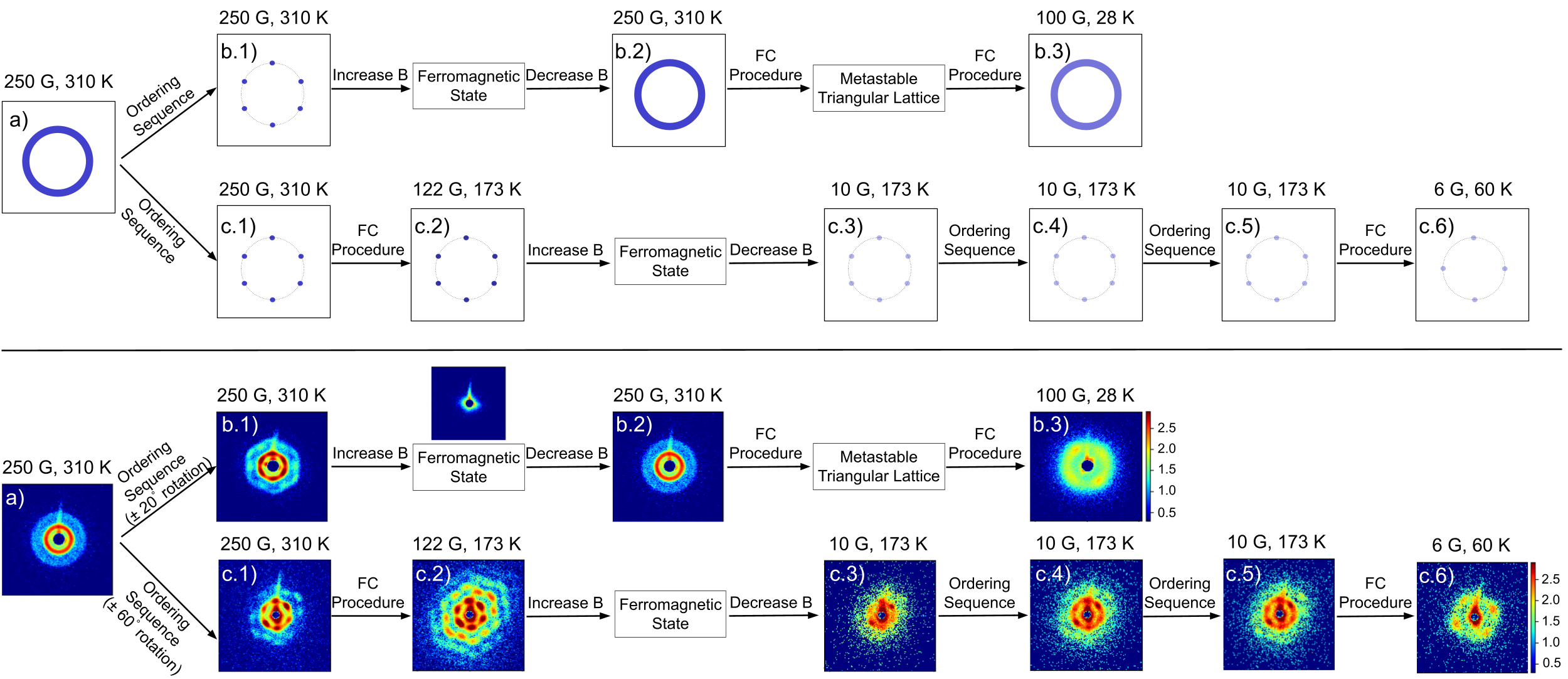}
\caption{ Schematic of expected SANS results for outlined experimental procedure (upper panel), and corresponding real SANS results (lower panel). Note: the central peak for the ferromagnetic phase is masked by the beam block, while prominent flares are able to leak out in the SANS image. All SANS images are normalized to a fixed number of standard monitor counts to enable direct comparison between images, with the same scale for the color plots. Note the colorbar is a log-scale of intensity. }\end{figure*}

The presence of secondary and tertiary scattering rings may provide additional information as to the long range magnetic order of the skyrmion lattice through reconstructions with higher-order diffraction peaks. Unfortunately, multiple scattering tends to overwhelm higher-order harmonics in the underlying structure. From an experimental point of view, Renninger scans may be employed to quantitatively distinguish the two mechanisms by ``rocking out" the condition for multiple scattering. Renninger scans were performed in a bulk sample of MnSi, revealing higher-order diffraction to arise from an interference effect \cite{adams2011long}. These scans may be performed in the future on this sample as a function of field and temperature to map out higher-order versus multiple scattering contributions in phase-space. Alternatively, in theory, given an ideal skyrmion sample with instrument resolution limited peak widths, second-order diffraction is indistinguishable from multiple scattering which takes the form of a self-convolution of the first-order peaks with themselves. However, in the case of a disordered sample, the two effects become distinguishable as the first-order diffraction pattern is convolved with some kernel which has cylindrical symmetry. As a result, the radial and angular peak profiles of the primary and secondary rings will be equivalent for the case of higher-order diffraction. For the case of multiple scattering, the 12 secondary peaks will have alternating larger radial widths and smaller angular widths relative to the 6 primary peaks. To determine the approximate ratio of the two effects one can, in theory, unwrap the images in polar coordinates and determine a linear combination of the simulated higher-order diffraction and multiple scattering images that would produce the experimentally measured intensities in frequency space. Unfortunately, given the low signal-to-noise ratio for the secondary peaks in these datasets, this type of analysis is not trivial.

In conclusion, we have shown skyrmion order, and ordering ability, to vary as a function of phase. Discrepancies in skyrmion lattice transition and metastable phenomena for ordered versus disordered samples were revealed through a newly established disordered-to-ordered square lattice transition pathway and metastable triangular lattice memory effect. These results greatly enhance our understanding of skyrmion stabilization and formation mechanisms for thermal versus metastable phases, demonstrating the interplay of topological stability with a tunable energy landscape of exchange, anisotropy, disorder (i.e. defects and pinning), field, and temperature. Ultimately, our work has provided a valuable account of skyrmion stabilization/formation and lattice restructuring dynamics as a function of disorder, establishing new pathways for skyrmion manipulation and enhanced stability for future devices.

\section{\label{sec:level4} Methods\protect\\}
 
\subsection{\label{sec:level1}Skyrmion ordering in the thermal equilibrium phase  \protect\\}

Unpolarized SANS was performed at the NG7-30m beamline at the National Institute for Standards and Technology (NIST) for a 15 m beam configuration and a neutron wavelength of 6 \AA (Fig. 1a) \cite{30mGlinka2018,kline2006reduction,disclaimer}. The sample used for these measurements was that of a polycrystalline cube of dimensions 3.4 mm x 3.3 mm x 3.0 mm, previously grown and characterized in \cite{henderson2020characterization}. Prior to changing any temperature or field parameters, the ground-state helical phase was verified at room temperature in zero field (Fig. 1b). A field-cooling procedure was executed from the ferromagnetic phase to enter the skyrmion envelope (Fig. 2), with the magnetic field applied along the neutron flight path (z-direction in Fig. 1a), lowered until 250 G, determined from the maximal dip in previous AC susceptibility measurements \cite{henderson2020characterization}. Using a skyrmion-ordering sequence \cite{dustin}, the sample was rotated symmetrically in the static magnetic field to precipitate an ordered and oriented skyrmion lattice (Fig. 1a).

\subsection{\label{sec:level2} Disordered metastable skyrmion transitions\protect\\}

To examine the influence of underlying disorder on skyrmion lattice transitions, the low temperature square lattice skyrmion phase (M-S-SkX) was investigated by way of the (D)E-T-SkX phase (Fig. 11a). The (D)E-T-SkX phase was regained from its ordered counterpart by saturation into the ferromagnetic phase, through the application of a strong magnetic field, and subsequent lowering of the field back into the skyrmion envelope (b.2 of Fig. 11). A square phase was then realized at 28 K in a field of 100 G (b.3 of Fig. 11).

\subsection{\label{sec:level3} Ordered metastable skyrmion transitions \protect\\}

In a separate SANS experiment, the (O)E-T-SkX phase was brought into (O)M-T-SkX phase via FC (c.2 of Fig. 11). In the (O)M-T-SkX phase, at a temperature of 173 K, the skyrmion phase was saturated into the ferromagnetic phase by increasing the field to 1 T, from its previous value of 10 G (indicated by the vertical arrow in Fig. 2). The field was then lowered back to 10 G (c.3 of Fig. 11) in an attempt to regain the phase to examine the nature of the metastability, and any potential memory effects. The ordering sequence was then performed in the resultant phase for two different field values (c.4 and c.5 of Fig. 11) in order to investigate the energetics and pinning dynamics of the persisting phase. Finally, the ordered square phase was entered (c.6 of Fig. 11) using another FC procedure for comparison with the disordered square phase, to understand the mechanisms of the transition.

\section{\label{sec:level5} Data Availability\protect\\}

Supporting data is available upon request from corresponding authors. 
 
\begin{acknowledgments}

This work was supported by the Canadian Excellence Research Chairs (CERC) program, the Natural Sciences and Engineering Council of Canada (NSERC) Discovery program, the Canada First Research Excellence Fund (CFREF), and the National Institute of Standards and Technology (NIST) Quantum Information Program. Access to SANS and CHRNS was provided by the Center for High Resolution Neutron Scattering, a partnership between the National Institute of Standards and Technology and the National Science Foundation under Agreement No. DMR-1508249. We would like to thank Jeff Krzywon and Tanya Dax for their assistance with SANS instrumentation, Steven Kline for his support regarding SANS fitting, and Dustin Gilbert for many helpful discussions.    
\end{acknowledgments}

\section{\label{sec:level7}Ethics declaration\protect\\}

The authors declare no competing financial or non-financial interests.

\nocite{*}
\typeout{}
\bibliography{refs}

\begin{thebibliography}{58}
\bibitem [{dis()}]{disclaimer}%
  \BibitemOpen
  \href@noop {} {\bibinfo {title} {Specific commercial equipment, instruments,
  or materials are identified in this paper to foster understanding. such
  identification does not imply recommendation or endorsement by the national
  institute of standards and technology, nor does it imply that the materials
  or equipment identified are necessarily the best available for the
  purpose.}}\BibitemShut {Stop}%
\end{thebibliography}

\end{document}